\documentclass[10pt,conference]{IEEEtran}
\IEEEoverridecommandlockouts

\usepackage{cite}
\usepackage{todonotes}
\usepackage{amsmath,amssymb,amsfonts}
\usepackage{algorithmic}
\usepackage{graphicx}
\usepackage{textcomp}
\usepackage{float}
\usepackage{xcolor}
\usepackage{multirow}
\usepackage{orcidlink}
\usepackage{hyperref}
\usepackage[moderate, tracking=normal]{savetrees}

\title{Approach Towards Semi-Automated Certification for Low Criticality ML-Enabled Airborne Applications}

\author{\IEEEauthorblockN{Chandrasekar Sridhar\textsuperscript{\textdagger}~\orcidlink{0009-0009-3679-6331}, Vyakhya Gupta\textsuperscript{\textdagger}~\orcidlink{0009-0003-0879-6226}, Prakhar Jain\textsuperscript{\textdagger}~\orcidlink{0009-0005-4221-2946}, Karthik Vaidhyanathan\textsuperscript{\textdagger}~\orcidlink{0000-0003-2317-6175}}
\IEEEauthorblockA{\textit{Software Engineering Research Center},  \textit{IIIT Hyderabad}, India\\
chandrasekar.s@research.iiit.ac.in, vyakhya.gupta@students.iiit.ac.in,
prakhar.jain@research.iiit.ac.in,\\
karthik.vaidhyanathan@iiit.ac.in
}
\thanks{\textsuperscript{\textdagger} All the authors contributed equally to this work.}
}

\begin{document}

\maketitle

\begin{abstract}
As Machine Learning (ML) makes its way into aviation, ML-enabled systems—including low-criticality systems—require a reliable certification process to ensure safety and performance. Traditional standards, like DO-178C, which are used for critical software in aviation, don’t fully cover the unique aspects of ML. This paper proposes a semi-automated certification approach, specifically for low-criticality ML systems, focusing on data and model validation, resilience assessment, and usability assurance while integrating manual and automated processes. Key aspects include structured classification to guide certification rigor on system attributes, an Assurance Profile that consolidates evaluation outcomes into a confidence measure the ML component, and methodologies for integrating human oversight into certification activities. Through a case study with a YOLOv8-based object detection system designed to classify military and civilian vehicles in real-time for reconnaissance and surveillance aircraft, we show how this approach supports the certification of ML systems in low-criticality airborne applications.

\end{abstract}


\section{Introduction}
\label{sec:intro}

The increasing integration of ML in aviation presents both opportunities and challenges, including applications with low criticality where flexibility and adaptability are highly valued but must still meet safety and reliability standards. Traditional certification standards such as DO-178C~\cite{178C} were developed for deterministic rule-based software, emphasizing strong traceability from requirements to code, a task that is often challenging in practice. However, ML systems (MLS) operate differently. They are inherently probabilistic and data-driven, meaning they learn from data rather than following predefined paths~\cite{tambon2022certify}. This fundamental shift in MLS' functionality has created gaps in existing certification standards~\cite{jenn2020identifying}, as these standards do not address the unique needs of ML, such as continuous model validation, data integrity, performance uncertainity handling, and more~\cite{validation,dataintegrity,caveness2020tensorflow,tambon2022certify,jenn2020identifying,bosch2021engineering,hullermeier2021aleatoric}.

The complex, dynamic nature of MLS introduce a major challenge in scaling certification processes. Manually certifying each component can be resource-intensive and prone to inconsistencies. Automation is therefore essential, not only to improve efficiency but also to ensure reliable, repeatable assessments that are crucial for ensuring the safety and reliability of ML-driven applications.
Recent guidelines from the European Union Aviation Safety Agency (EASA)~\cite{easapaper} highlight these gaps, recognizing that existing standards are insufficient for assessing ML behaviour across diverse operational scenarios and addressing issues like data drift. These limitations hinder the deployment of ML technologies in airborne systems, when maintaining both safety and reliability is critical. 

This paper addresses these challenges by proposing a semi-automated certification approach designed for low-criticality MLS in aviation. 
In this context, "low-criticality" corresponds to DO-178C Level D, corresponds to DO-178C Level D, where software failures result in minor failure conditions for the aircraft.
Our central focus of this research is to explore approaches for maintaining the validity of an MLS' certification throughout its operational life. Unlike traditional software, MLS may require periodic re-evaluation to maintain certification standards as the system continues to operate. This paper makes the first step in addressing these requirements by proposing a certification process for MLS. 
We present a structured classification approach for MLS to tailor verification extensiveness, an \textit{Assurance Profile} to quantify confidence, and methods for incorporating human oversight into certifying MLS in Level D criticality aviation.
\begin{figure*}[htbp]
\centerline{\includegraphics[scale=0.15]{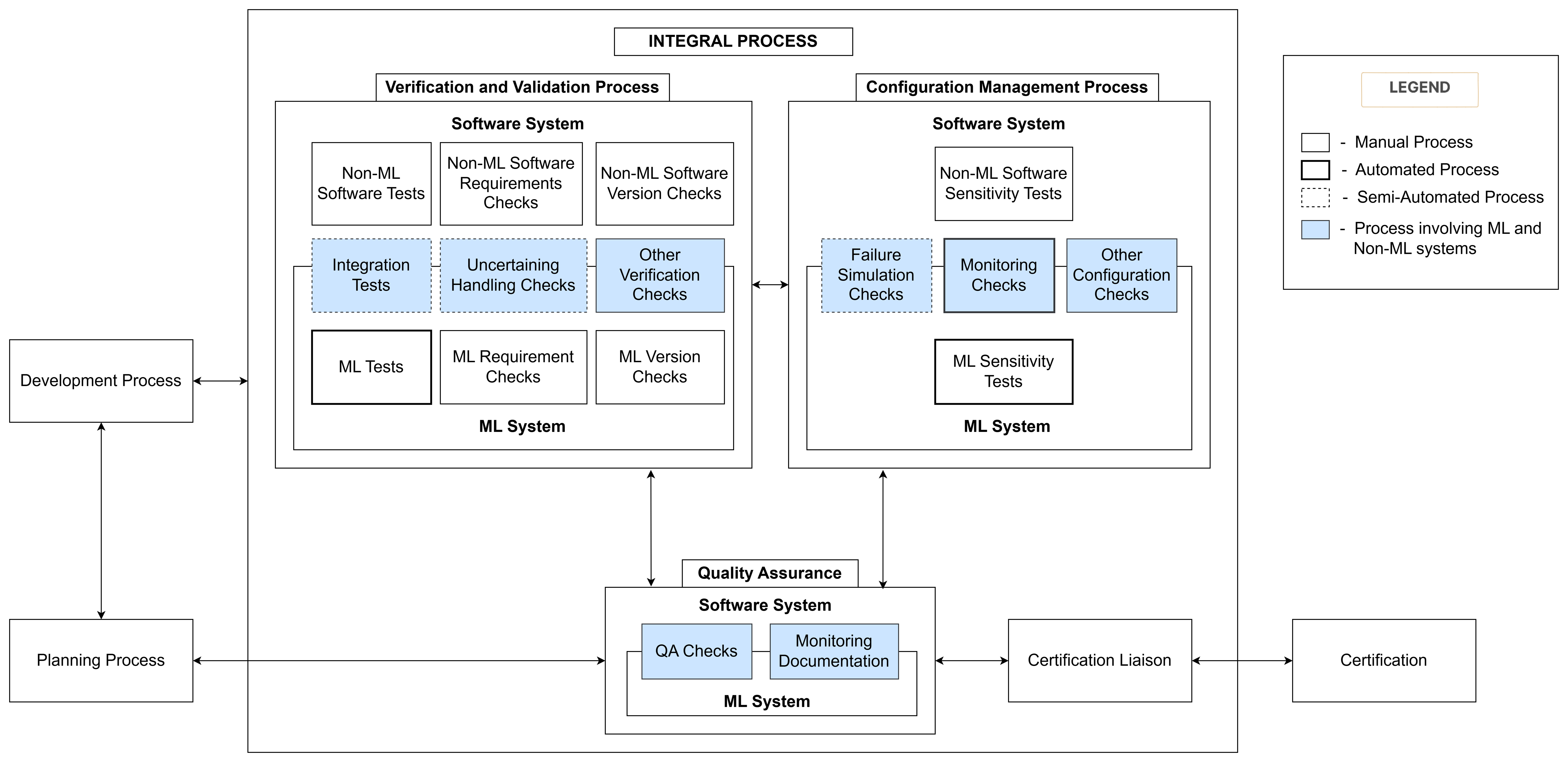}}
\caption{Proposed Integrated Certification Process for ML and Non-ML System}
\label{fig:semi_auto_approach}
\end{figure*}
To effectively address the mentioned challenges, we’re making a few key assumptions, similar to those mentioned by in~\cite{certairborne}. First, our focus is solely on the ML components of the system, while traditional, non-ML components 
are assumed to be certifiable under DO-178C without modification. Second, we are not addressing adaptive MLS that continue to learn and evolve while in operation, as these introduce further complexities. Our analysis is limited to systems where ML models are trained offline and remain static during operation.

To illustrate this approach, we examine a case study involving a YOLOv8-based object detection system~\cite{yolov8} , aimed at real-time classification in surveillance aircraft. 
The paper is structured as follows: Section~\ref{sec:related_work} discusses related work and challenges in certifying ML for aviation. Section~\ref{sec:case_study} describes the overview of the Air Sight case study. Section~\ref{sec:semi_automated_cert} presents the semi-automated certification approach. Section~\ref{sec:prelim_eval} provides preliminary evaluation, and Section~\ref{sec:conclusion} concludes with insights and future directions.

\section{Related Work}
\label{sec:related_work}

This work builds on previous research in MLS certification for low-criticality airborne applications, particularly Dmitriev et al.\cite{certairborne}, who address MLS-specific certification challenges in aviation. Sculley et al.\cite{hiddentechdebt} further highlight how poor design decisions can increase maintenance costs and degrade performance, reinforcing the need for reliable design and testing practices.
Tambon et al.\cite{tambon2022certify} review certification strategies for safety-critical ML systems, identifying gaps related to MLS’s probabilistic nature and the need for continuous validation and interpretability—factors we address with our semi-automated approach. Similarly, Pérez-Cerrolaza et al.\cite{industrialtransport} emphasize specialized certification methods for AI in safety-critical settings, highlighting the varied complexities of certifying MLS across sectors. ~\cite{yang2022correctnessverificationneuralnetworks,Pei_2017,tian2018deeptestautomatedtestingdeepneuralnetworkdriven} talk about testing and explaining deep models. For reliable MLS certification, we draw inspiration from the ML Test Score ~\cite{themltest}, structuring test cases and scoring to address challenges. This approach focuses on Level D criticality MLS certification, also adapting principles from critical space systems~\cite{aimlspace} that emphasize traceability and validation for aviation needs.
Additionally, Dmitriev et al.\cite{toolqualification} explore the limitations of tool standards like DO-330\cite{330} and DO-200B~\cite{200B} when applied to MLS, recommending adjustments to support MLS-specific workflows, including data management and model training. We also incorporate structured certification methods from Delseny et al.~\cite{whitepaper}, which emphasize safety and reliability for MLS. 
This work also draws insights from recent research on best practices and standards proposed for ML-enabled systems and cyber-physical systems, which address challenges in ensuring robustness, systematic testing, and compliance throughout their lifecycle~\cite{bello2024towards,gariel2023framework,rierson2017developing,sivakumar2024came,chandrasekaran2023test,anisetti2023rethinking,unknown}. Together, these insights support a certification process that addresses the challenges of MLS while maintaining standards for Level D criticality applications.

\section{Air Sight - Case Study} 
\label{sec:case_study}
This case study examines the certification process for a machine learning-based object detection system designed for low criticality airborne applications in a military context. The system, which utilizes a YOLOv8 model trained on a diverse dataset of military and civilian vehicles, aims to classify various types of air and ground assets in real-time.  Intended for integration into reconnaissance and surveillance aircraft, the system processes visual data to identify and classify vehicles of interest, enhancing situational awareness for pilots and operators. While not directly involved in flight-critical operations, the system's outputs may influence tactical decision-making, highlighting the need for a robust certification process.
This serves as a foundation for exploring how existing software certification standards, particularly DO-178C, can be extended to accommodate the unique characteristics of ML-based systems in low criticality airborne applications.
The requirements for this certification approach were partially shaped by guidance from stakeholders specialized in the airborne system domain. Their expertise served as a foundation for exploring how existing software certification standards, particularly DO-178C, can be extended to accommodate the unique characteristics of ML-based systems in Level D airborne applications.


\section{Semi-Automated Certification Approach}
\label{sec:semi_automated_cert}

The semi-automated certification approach, as illustrated in Figure \ref{fig:semi_auto_approach}, addresses two primary challenges: {\em validating model consistency} and {\em assessing overall system resilience}. This approach integrates both manual and automated processes across multiple stages, including Verification, Configuration Management, and Quality Assurance, which are exemplified through the Air Sight case study in the following sections.

\subsection{Classification Approach for ML-enabled Systems}
\label{subsec:classificationapproach}

The approach uses a multi-axial classification which defines the certification requirements for MLS in aviation, tailoring rigor to system attributes. It is defined as \( C = \langle c_{\text{crit}}, c_{\text{aut}}, c_{\text{model}}\rangle \) where: \textbf{\textit{a) System Criticality ($c_{\textit{crit}}$):}} follows DO-178C guidelines, ranging from Level A (highest criticality) to Level E (no impact on safety). This study focuses on Level D, representing low-criticality applications that still require basic safety and reliability standards. \textbf{\textit{b) Autonomy Level ($c_{\textit{aut}}$):}} assessed based on EASA’s levels~\cite{easapaper}, spanning from Level 1A (high human supervision) to Level 3B (full autonomy). Higher levels warrant stricter certification due to reduced human oversight. \textbf{\textit{c) ML Complexity ($c_{\textit{model}}$):}} Three levels adjust certification rigor based on model sophistication:
\textbf{\textit{i) Level 1}}: Simple models (e.g., linear or logistic regression) with high interpretability; requires minimal validation \( \langle V_1 \rangle\).
\textbf{\textit{ii) Level 2}}: Intermediate models (e.g., decision trees, random forests) with moderate adaptability; needs additional validation for reliability \( \langle V_2 \rangle\).
\textbf{\textit{iii) Level 3}}: Complex models (e.g., neural networks, CNNs) with nonlinear, high-dimensional decision boundaries; demands extensive validation due to challenges in interpretability \( \langle V_3 \rangle \).

Air Sight is classified as \( C = \langle c_{\text{crit}}: D, c_{\text{aut}}: 2A, c_{\text{model}}: 3\rangle \), meaning it operates at a low-criticality level, with moderate autonomy requiring human oversight, and utilizes a complex ML model like YOLOv8. This classification ensures the certification process is tailored to focus on appropriate levels of validation, robustness, and interpretability for safe and reliable operation.

\subsection{Certification Layers}
\label{subsec:certification_layers}

The proposed approach comprises of three layers, each addressing distinct assurance aspects for ML-enabled systems. These layers are based on the system's classification, \( C\). \textbf{\textit{i) Base Layer (DO-178C Processes):}} covers requirements capture, software development, configuration management, verification, and integration. This layer establishes a solid foundation for certifying the non-ML components of \( c_{\text{crit}} = D  \) system. \textbf{\textit{ii) ML Layer (ML-Specific Assurance):}} focuses on ML-specific assurance, encompassing the entire ML lifecycle from dataset preparation and model training to hyperparameter optimization and performance evaluation. For \( c_{\text{model}} = 3 \), the processes are enhanced to address increased non-linearity and dimensionality. \textbf{\textit{iii) Human Factors Layer (Human-AI Interaction Assurance):}} targets the usability, transparency, and reliability of AI components in supporting operator decision-making. For \( c_{\text{aut}} = 2A \) in Air Sight, we evaluate how the system enhances trust and effectiveness through usability testing in simulated environments, ensuring it improves situational awareness and reduces cognitive workload. The Base Layer follows standard practices, and the focus here is on the ML and Human Factors Layers, which apply specific methodologies to meet operational and safety requirements.

\subsection{Certification Process}
\label{subsec:certification_process}

Automation is key to scalable, consistent evaluations in this certification approach. Solely manual certification is impractical for MLS due to model complexity and the vastness of data. Automating checks like data validation, performance monitoring, and robustness testing ensures reliable, ongoing validation. It also supports continuous compliance by detecting shifts in data relevance or model performance over time.


Central to our approach is the \textit{Assurance Profile}, a transparent summary of compliance for the ML layer. It uses a structured scoring system to derive a final certification score, \( S_{\text{cert}} \), based on both manual reviews and automated checks. This score translates into a Confidence Level (\( \sigma \)) offering an intuitive assurance metric for stakeholders.

As illustrated in Figure \ref{fig:semi_auto_approach}, the approach includes key processes: Development (Dev), Verification and Validation (V\&V), Quality Assurance (QA), and Configuration Management (CM). After each process, artifacts and evidence—such as Dataset Documentation, Model Configuration Reports, verification results, and monitoring logs—are generated to support the certification. Each process, denoted as \( \text{Process}_i = \{\text{Dev}, \text{V\&V}, \text{QA}, \text{CM}\} \), is comprised various activities \( \{\text{act}_1, \text{act}_2, \ldots\} \), each evaluated through a series of manual and automated checks. The scores from manual evaluations, along with the percentage of automated checks passed, are aggregated to assess the performance each activity by producing $S_{\text{act}_i}$.

For each process, the score $S_{\text{process}}$ is the weighted sum of its activities:
\[
S_{\text{process}} = \sum_{i=1}^{n} w_{\text{act}_i} \cdot S_{\text{act}_i}
\]
The weights for each activity, \( w_{\text{act}_i} \), are determined as a factor of the system's classification \( C \) and contextual factors, including activity complexity and its impact on system performance, where \( S_{\text{act}_i} \) is the score of the \( i \)-th activity.

The final certification score \( S \) aggregates the scores from all processes, with each process weighted by \( w_{\text{process}_j} \), which is similarly a factor of \( C \) and contextual influences:
\[
S = \sum_{j=1}^{4} w_{\text{process}_j} \cdot S_{\text{process}_j}
\]

The certification confidence level \( \sigma(S) \) categorizes the MLS score \( S \in [0, 100] \) into five confidence levels:

\[
\sigma(S) = 
\begin{cases} 
    \text{Optimal Assurance} & : 90 \leq S \leq 100 \\
    \text{Strong Assurance} & : 80 \leq S < 90 \\
    \text{Moderate Assurance} & : 70 \leq S < 80 \\
    \text{Limited Assurance} & : 60 \leq S < 70 \\
    \text{Insufficient Assurance} & : S < 60 
\end{cases}
\]
\vspace{0.05cm}

Where: \textbf{Optimal Assurance} indicates maximum reliability; \textbf{Strong Assurance} suggests high performance with minor areas for improvement; \textbf{Moderate Assurance} meets criteria, with periodic review advised; \textbf{Limited Assurance} implies marginal reliability, requiring closer oversight; and \textbf{Insufficient Assurance} denotes significant deficiencies needing correction.

\subsubsection{Planning Process}
\label{subsec:planning_process}
The planning process, as illustrated in Figure~\ref{fig:semi_auto_approach}, defines the foundational requirements for functional and non-functional aspects, forming the basis for traceability of the ML Layer. It establishes criteria for dataset diversity, label accuracy, and model performance, setting thresholds for key assurance checks. An uncertainty management strategy is included, setting confidence levels to handle ambiguous cases and guide human oversight.

For the Air Sight system, these requirements classify it as \( C = \langle c_{\text{crit}}: D, c_{\text{aut}}: 2A, c_{\text{model}}: 3 \rangle \). This $C$, along with detailed requirements documentation, informs the weights assigned in the \textit{Assurance Profile}. It also guides the core processes outlined in Figure~\ref{fig:semi_auto_approach}, so that planning aligns with each layer's evaluation criteria.


\subsubsection{Development Process}
\label{subsec:development_process}

This process (refer Figure~\ref{fig:semi_auto_approach}), focuses on building the ML dataset, model, and non-ML components, adhering to predefined requirements using automated tools for data integrity and model configuration. These tools verify dataset completeness, label accuracy, and class balance, while model architecture and hyperparameters are documented. Training processes must align with requirements. Unlike traditional software, MLS require thorough certification of the development activities for model reliability. This is less emphasized in lower-criticality systems but essential for robust validation. Integration documentation covers how ML and non-ML components interact, including interface specifications, data flow mappings, and system-level integration.

For Air Sight, the dataset is assessed for indicators like environmental diversity, which affects model adaptability. Some metadata lacks specific details on lighting or weather, lowering dataset score, but thorough preprocessing and augmentation like rotation and scaling improve the model’s ability to detect objects under different perspectives. Integration documentation details how augmented data flows into the system's tactical display, ensuring real-time situational awareness without impacting operational safety. In the context of 
Level D criticality, it allows flexibility in dataset limitations, while Level 3 model complexity requires fine-tuning and detailed documentation of model adjustments.


\subsubsection{Verification and Validation Process}
\label{subsec:verification_and_validation}

This process, as shown in Figure~\ref{fig:semi_auto_approach}, is essential for building the \textit{Assurance Profile}, blending traditional V\&V processes with methods tailored for ML components. This process evaluates the ML Layer and ensures that it meets the required aviation safety and reliability standards. It includes the following key activities:

\begin{figure}[htbp]
\centerline{\includegraphics[width=\linewidth]{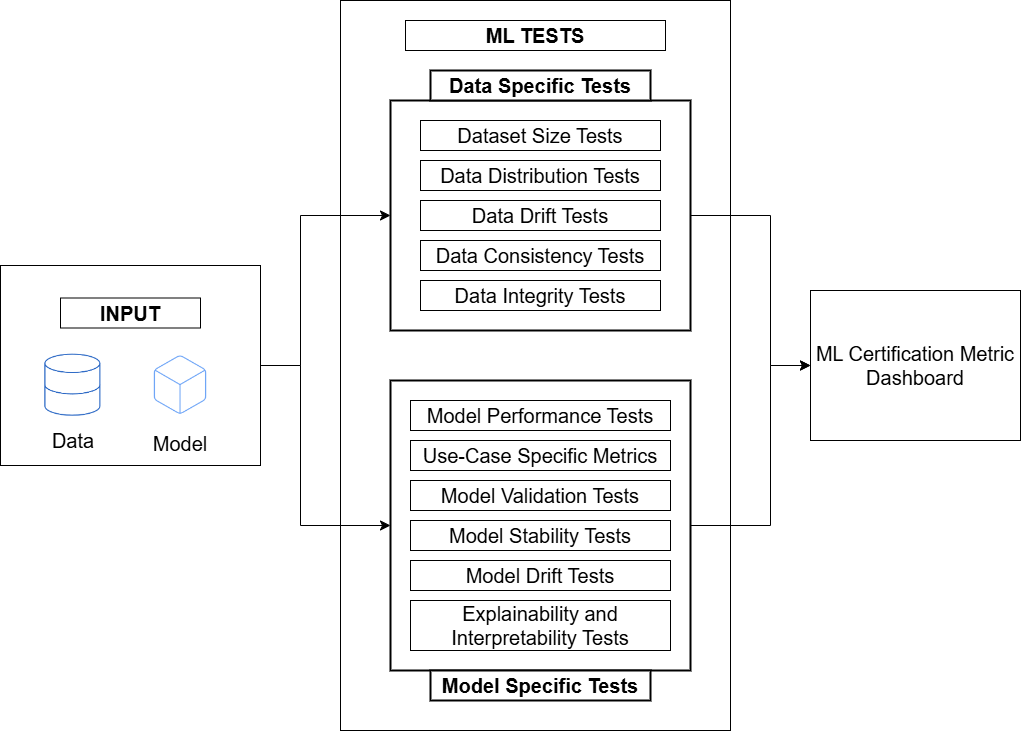}}
\caption{Automated ML Testing}
\label{fig:ml_tests}
\end{figure}
\textbf{\textit{a) Automated Processes:}}
Figure \ref{fig:ml_tests} shows our ML certification testing approach, covering key datasets and model checks for automated testing.
Our ML certification approach includes automated testing to ensure reliability and robustness. 
\textbf{\textit{i) Model Performance}} is evaluated using metrics like precision and F1-score across varied operational scenarios to validate real-time detection capabilities. 
\textbf{\textit{ii) Robustness Testing}} examines the model's performance under real-world edge cases such as occlusion, motion blur, and noise, ensuring functionality without adversarial scenarios. \textbf{\textit{iii) Dataset Certification}} checks for integrity, distribution, and anomalies like data drift to ensure the dataset reflects the operational environment, minimizing biases or performance gaps.

\textbf{\textit{b) Semi-Automated Processes:}} These focus on \textbf{\textit{i) System Integration}} to 
verify seamless communication between ML and non-ML components across interfaces and data flows. \textbf{\textit{ii) Uncertainty Handling}} assesses the system’s response to ambiguous cases, requiring human oversight for complex outputs and timely intervention. Due to the non-deterministic nature of AI, it is crucial to set acceptable error tolerances for expected scenarios. Additionally, fail-safe mechanisms must be in place to address unexpected or unsafe behavior, mitigating risks in line with aviation standards. \textbf{\textit{iii) Human Factors}} testing incorporates usability assessments through simulated decision-making environments and operator feedback, ensuring the system aligns with user needs.

For Air Sight, key model metrics like precision, recall, mAP, accuracy were as per requirements, despite some class performance imbalance. Testing revealed degradation under conditions like noise and varying environmental factors. Dataset certification identified image property outliers, such as unusual lighting, which could affect model reliability. With $c_{\text{model}}$=3, extensive validation, \( \langle V_3 \rangle \) was performed.

\subsubsection{Quality Assurance}
\label{subsec:quality_assurance}

As depicted in Figure~\ref{fig:semi_auto_approach}, in a Level D MLS, QA ensures compliance with DO-178C standards across all certification layers, focusing on consistency and stability. It integrates automated and human-supervised tools to verify adherence to the Base Layer, ML Layer, and Human Factors Layer, ensuring the reliability of both software components and ML models. Key QA activities for Level D systems include:
\textbf{\textit{i) Post-Certification Operational Monitoring:}} QA ensures dataset relevance, model integrity, and ML accuracy through testing for robustness and interpretability throughout operational use. A structured monitoring plan tracks performance and user feedback, with recertification triggers for significant updates to maintain compliance.
\textbf{\textit{ii) Adherence Audits \& Documentation Reviews:}} Routine audits confirm that development, testing, and human factors assessments meet standards, covering ML training, dataset management, and interface design. Formal reviews ensure proper documentation of model versions, datasets, and changes for traceability and compliance.
\textbf{\textit{iii) Usability Assessment:}} Usability evaluations verify that human-AI interactions enhance situational awareness and reduce cognitive load, supporting effective operator decision-making.

In the Air Sight system, QA activities identified key areas for improvement, particularly in post-certification monitoring and usability. Low usability scores stemmed from difficulties in interpreting AI decisions during complex, fast-changing scenarios. Post-certification monitoring also revealed label drift due to the system encountering previously unrepresented vehicles or conditions. Despite these challenges, the QA framework ensures ongoing compliance and provides insights for improving model performance and user interaction.

\subsubsection{Configuration Management Process}
\label{subsec:config_management}

CM process (refer Figure~\ref{fig:semi_auto_approach}) for Level D MLS ensures consistency across all components, including the user interface, ML model, and datasets. By combining automated and human-supervised activities, the CM process focuses on managing significant updates that impact performance and human interaction. Key CM activities include:

\textbf{\textit{i) Configuration Identification and Version Control:}} All components, including ML models, datasets, and human-interface elements, are identified and documented to create a stable baseline. Major updates, such as model retraining or interface adjustments, undergo appropraite version control.
\textbf{\textit{ii) Configuration Audits and Compliance:}} Regular audits verify that all changes align with system standards and do not negatively affect the user experience or system performance. These checks prevent updates from disrupting operator situational awareness or increasing the complexity of human-AI interactions.
\textbf{\textit{iii) Post-Certification Update Management:}} Procedures exist for updates to both the ML model and user interface after certification. Significant updates follow steps for documentation, review, and a potential recertification.

For Air Sight, these processes are essential for managing updates to the model, datasets, and user interface components. While major changes to YOLOv8 and the dataset are properly version-controlled, the current process lacks a structured approach for post-certification updates due to the experimental nature of the system. This gap in update management impacts the overall score in this domain.


\section{Preliminary Evaluation}
\label{sec:prelim_eval}


The YOLOv8 model, used for object detection in the Air Sight case study, was fine-tuned on a Nvidia GTX 1080 Ti with 12 GB VRAM using the~\cite{mil-dataset} dataset, ensuring that it could accurately classify various types of air and ground assets. Our approach to certify the system was partially realized using Deepchecks~\cite{chorev2022deepchecks} on a Ubuntu 22.04 machine with 16GB RAM powered by 4.7 GHz 12 Gen Intel i7 processor. Deepchecks was used specifically to assess various types of drifts.  Table \ref{tab:final_profile} offers a comprehensive summary of the Air Sight system’s certification journey, synthesizing scores from automated, semi-automated, and manual reviews across critical assurance dimensions. Through our semi-automated certification approach\footnote{\href{https://github.com/sa4s-serc/MLCert}{https://github.com/sa4s-serc/MLCert}}, the example below illustrates the Air Sight system's current status against the outlined criteria for a Level D classification.


\begin{table}[h!]
\caption{Assurance Profile for Air Sight System}

\centering
\resizebox{\linewidth}{!}{
\begin{tabular}{|l|c|c|c|}
\hline
\multicolumn{4}{|c|}{\textbf{Air Sight System Certification Summary (DO-178C Level D)}} \\ \hline
\textbf{Process} & \textbf{Score (100)} & \textbf{Weight} & \textbf{Weighted Score} \\ \hline

Dataset Quality & 80.0 & 0.40 & 32.0 \\ 
Model Documentation & 82.0 & 0.35 & 28.7 \\ 
Integration Documentation & 70.0 & 0.25 & 17.5 \\ \hline
\textbf{Total Development Score} & \textbf{} & \textbf{} & \textbf{78.2} \\ \hline

\multicolumn{4}{|c|}{\textbf{2. Verification \& Validation (V\&V)}} \\ \hline
Model Performance & 95.0 & 0.25 & 23.8 \\ 
Robustness Testing & 92.0 & 0.25 & 23.0 \\ 
Dataset Certification & 84.8 & 0.20 & 17.0 \\ 
System Integration & 88.0 & 0.15 & 13.2 \\ 
Human Factors & 87.0 & 0.15 & 13.0 \\ \hline
\textbf{Total V\&V Score} & \textbf{} & \textbf{} & \textbf{90.0} \\ \hline

\multicolumn{4}{|c|}{\textbf{3. Quality Assurance (QA)}} \\ \hline
Post-Certification Monitoring & 55.0 & 0.35 & 19.3 \\ 
Usability Assessment & 52.0 & 0.35 & 18.2 \\ 
Audits and Reviews & 50.0 & 0.30 & 15.0 \\ \hline
\textbf{Total QA Score} & \textbf{} & \textbf{} & \textbf{52.5} \\ \hline

\multicolumn{4}{|c|}{\textbf{4. Configuration Management (CM)}} \\ \hline
Version Control & 65.0 & 0.40 & 26.0 \\ 
Configuration Identification & 60.0 & 0.35 & 21.0 \\ 
Change Management & 58.0 & 0.25 & 14.5 \\ \hline
\textbf{Total SCM Score} & \textbf{} & \textbf{} & \textbf{61.5} \\ \hline

\multicolumn{4}{|c|}{\textbf{Final Certification Summary}} \\ \hline
Development & 78.2 & 0.30 & 23.5 \\ 
V\&V & 90.0 & 0.35 & 31.5 \\ 
QA & 52.5 & 0.20 & 10.5 \\ 
SCM & 61.5 & 0.15 & 9.2 \\ \hline
\textbf{Final Certification Score} & \textbf{} & \textbf{} & \textbf{74.7} \\ \hline

\end{tabular}
}
\label{tab:final_profile}
\end{table}





\subsection{Certification Details}



The \textit{Assurance Profile} for the Air Sight system presents a detailed evaluation of its certification readiness, resulting in a Final Assurance Score of 74.7\%, indicating a 
Moderate Assurance level as per the scoring benchmark \ref{subsec:certification_process}. Strong results in the V\&V and Development stages reflect the system's robust dataset quality, clear model documentation, and effective validation processes. However, the lower scores in QA and CM suggest areas that need improvement, particularly in post-deployment monitoring and version control.

This initial certification assessment demonstrates that Air Sight meets the compliance criteria for DO-178C Level D criticality, suitable for applications with lower safety risks. Given its classification $C$, full recertification is not necessary within a typical operational lifecycle. Instead, Air Sight will undergo targeted drift checks, which are periodic evaluations to ensure the model’s performance remains consistent over time, especially in response to changes in data patterns or operational environments. This avoids the burden of a full certification process while maintaining a focus on detecting significant shifts early.

The need for improved QA signals the importance of strengthening ongoing audit processes to ensure sustained reliability. With the system's autonomy and ML complexity levels, these- coupled with frequent drift checks— will help proactively manage any changes, ensuring that Air Sight continues to meet its operational requirements effectively and safely.

\subsection{Recertification Triggers}

To ensure the continued safety and reliability of the Air Sight system, effective recertification planning is crucial. The conditions to prompt this, as outlined in CM and QA in Section~\ref{subsec:certification_process}, are:
\textbf{\textit{i) Performance Degradation:}} Recertification is required if operational accuracy drops below desired threshold.
\textbf{\textit{ii) Dataset Shift:}} A significant change (over 30\%) in dataset distribution or model architecture triggers recertification to evaluate data drift or model evolution.
\textbf{\textit{iii) Environmental Changes:}} Major updates to operational conditions, such as new deployment scenarios or mission requirements.
\textbf{\textit{iv) Updates to Uncertainty Handling:}} Changes to uncertainty management strategies" revised confidence thresholds or new failure modes.

\subsection{Evaluation of our Certification Approach}

To assess the validity of our approach, we introduced gaussian noise into the validation set derived from the~\cite{mil-dataset} dataset. This simulated real-world conditions where sensor or environmental inconsistencies may degrade data quality. Despite these perturbations, the MLS achieved the performance metrics: 79\% precision and 75\% recall and a mAP score of 81\%. Compared to the unperturbed dataset, the performance degradation was minimal, reinforcing the MLS's robustness under real-world conditions. While prediction drift was observed during testing, it remained within the acceptable threshold for Level D criticality systems ($<$30\%), negating the need for recertification. The evaluation confirmed that the ML component of the system maintains acceptable performance and compliance with certification criteria under perturbed conditions, validating its readiness for deployment in Level D applications.




\section{Conclusion}
\label{sec:conclusion}
 The integration of ML-based systems into aviation introduces new certification challenges, as illustrated by the Air Sight case study. Traditional standards like DO-178C lack the adaptability needed for certifying data-driven, evolving ML systems. Our semi-automated certification approach addresses these gaps by focusing on continuous model performance evaluation, data quality management, and resilience in varied operational contexts. We also introduce an \textit{Assurance Profile} for ML systems, akin to a nutrition label. This \textit{Assurance Profile} provides a structured way forward in certifying ML systems, capturing the nuanced characteristics of ML systems. 

We plan to extend our approach to accommodate a broader range of use cases, enabling its application to more complex ML systems across various domains as well as for ML systems with higher criticality levels.

\bibliographystyle{ieeetr}
\bibliography{main}

\end{document}